# A Compact Low-Cost Low-Maintenance Open Architecture Mask Aligner for Fabrication of Multilayer Microfluidics Devices

Quang Long Pham,[a] Nhat-Anh N. Tong,[a] Austin Mathew,[b] Roman S. Voronov[a,†]

[a]Otto H. York Department of Chemical, Biological and Pharmaceutical Engineering, New Jersey Institute of Technology, Newark, NJ 07102, USA
[b]Department of Biomedical Engineering, New Jersey Institute of Technology, Newark, NJ 07102, USA.
[†]Corresponding author. Email: *rvoronov@njit.edu*

ABSTRACT:
A custom-built mask aligner (CBMA), which fundamentally covers all the key features of a commercial mask aligner, while being low cost, light weight, and having low power consumption and high accuracy is constructed. The CBMA is comprised of a custom high fidelity LED light source, vacuum chuck and mask holder, high-precision translation and rotation stages, and high resolution digital microscopes. The total cost of the system is under $7,500, which is over ten times cheaper than a comparable commercial system. It produces a collimated ultraviolet illumination of 1.8-2.0 mW cm$^{-2}$ over an area of a standard 4-inch wafer, at the plane of the photoresist exposure; and the alignment accuracy is characterized to be < 3 µm, which is sufficient for most microfluidic applications. Moreover, this manuscript provides detailed descriptions of the procedures needed to fabricate multilayered master molds using our CBMA. Finally, the capabilities of the CBMA are demonstrated by fabricating two and three-layer masters for micro-scale devices, commonly encountered in biomicrofluidics applications. The former is a flow-free chemical gradient generator and the latter is an addressable microfluidic stencil. Scanning electron microscopy is used to confirm that the master molds contain the intended features of different heights.

*Keywords*: mask aligner, photolithography, microfluidics, microfabrication, multi-height

## 1. Introduction
Multilayer photolithography has played a central role in the microfabrication of multi-height photoresist master molds for polydimethylsiloxane (PDMS) microfluidic devices, which is widely used in various fields of biological, micro electro-mechanical systems (MEMs), micro total analysis systems (µTAS), sensors and other applications.[1-6] Moreover, with the current explosion of the lab-on-a-chip technology in diagnosis and fundamental medical research, the need for the miniaturization of systems via deposition and etching procedures is greater than ever before.[7-9] However, the high cost of the equipment and its maintenance, combined with the expensive and space-consuming installation requirements associated with commercial systems, limit the access of many small laboratories and companies to manufacturing microfluidics devices in-house. For example, at the center of the photolithography fabrications is a mask aligner, which simultaneously provides precise mask-to-wafer alignment and generates a uniform ultra violet (UV) illumination over an exposed photoresist surface. However, current commercial systems (i.e. Karl Suss, OAI, and EVG) are very costly (e.g., $60,000-100,000 capital investment, plus $1,000-3,000 annual maintenance), heavy (> 250 lbs.), and relatively bulky (> 700 in$^2$). Therefore, there is an unmet need for photolithography systems that are fully functional, yet compact and affordable. To that end, this manuscript aims to present a custom-built mask aligner (CBMA), that can be constructed easily and for as little as $7,500 (and potentially even lower, if an in-house machine shop is used).

A collimated UV light source is the key component of any mask aligner, as it is directly responsible for creating the master molds for the microfluidics devices by crosslinking photoresist coated on wafers in a masked pattern. Typically, a high-pressure mercury lamp is used for this purpose by commercial systems. However, the UV light alone requires a significant capital investment (~$10,000-20,000), and accounts for a dominant portion of a mask aligner's annual maintenance costs. Moreover, the mercury lamps tend to provide a broadband illumination (bad for fabricating tall features) with an intensity that often drifts with time. Consequently, several efforts have been made to produce low-cost alternatives to the mercury lamp. For example, Huntingtan et al[10] introduced a portable light source composed of an array of hundreds of light emitting diodes (LEDs), which could be powered by AA batteries. Though such sources can provide a relatively uniform illumination that yields sub-micron features, they are not suitable for thick photoresist masters (i.e. over 100 µm) due to the large divergence of the illumination. However, many microfluidics applications require features this high. Consequently, a more recent light source addressed this problem by improving the light collimation using optical lenses that help to reduce the divergence of the light beam.[11] As a result, it was able to pattern thick masters of up to several hundred microns, while using an array of just 9 high power 365nm LEDs. Therefore, we chose this light source for our CBMA for its superior quality and minimalistic design.

In addition to the light source, a complete mask aligner system should have the capability for a mask-to-wafer alignment, which is critical for multi-layer soft lithography. Alignment systems have been made by other researchers. For example, Li et al[12] introduced a desktop aligner that combines digital microscopes with high-precision translation stages for aligning polydimethylsiloxane (PDMS) slabs. However, photolithography is not the main purpose of this system, and thus it is not equipped with a UV light source for photoresist exposure. Therefore, there is a need for an alignment system with an integrated high fidelity light source, capable of deep (i.e. > 100 µm) photoresist exposure.

To that end, we designed a low cost alternative to current commercial systems that is well equipped with the fundamental functions of a complete mask aligner: a uniform UV light with deep photoresist penetration, high-resolution rotation and translation stages, vacuum chucks and mask holder, and digital microscopes for high-precision alignment, etc. Along with the instructions for how to build the system, this manuscript also provides the detailed procedures for how to optimize its performance and manufacture multi-layer master molds with a high accuracy. Given that this system is affordable and low maintenance, it is expected that it will have widespread applications, especially in small and medium laboratories that do not have access to an in-house microfabrication facility.

## 2. Key Features & System Description

Key features of our CBMA include cost-savings, light weight, space savings, and an alignment accuracy sufficient for most typical microfluidics applications. A comparison of some of these the key features between the CBMA and a typical commercial system is given in **Table 1**. Relative to a commercial system (i.e. Karl Suss MA100, EVG 620, OAI 200), our system is an order of magnitude cheaper, weighs ~5 times less, occupies ~5 times less space, and consumes ~50 times less power. Moreover, it has an open adjustable architecture and a negligible annual maintenance cost. Finally, although the resolution of CBMA is not as high as that of a commercial system, it is still sufficient for most MEMS and microfluidics applications.

**Table 1**. Key feature comparison between the CBMA and a typical commercial mask aligner.

| Features | Custom-built | Commercial |
|---|---|---|
| Cost/Price, *$* | < 7,500 | 60,000 - 120,000 |
| Weight, *lbs* | < 50 | > 250 |
| Light source power, *W* | ~ 22 | > 1000 |
| Width x Depth, *in$^2$* | 15 x 10 | 32x23 |
| XY Alignment Accuracy, *µm* | 2.96 | < 0.5 |

**Table 2.** Cost breakdown showing the approximate costs of the main components of our CBMA.

| Item | Part Numbers (quantity) | Company | Cost ($) |
|---|---|---|---|
| Translation & rotation stages | 423-MIC (3), 461-XZ-M(1) 360-90 (2), UTR80 (1), 39 (1) | Newport, Irvine, CA | 4,100 |
| Machined adapters | - | Zera Development, S. Clara, CA | 1,000 |
| Digital microscopes | AD4113T (2) | Dino-lite, Torrance, CA | 900 |
| UV LED | UV LED 897-LZ110U600 (9) | Mouser, Mansfield, TX | 390 |
| Machined wafer chuck | - | Emachineshop, Mahwah, NJ | 300 |
| Dovetail optical Rail & carrier | 10R300 (2), 20C (2) | Optic Focus Solution, China | 280 |
| UV-collecting lens | FCN12592_LE1-D-COP (9) | Mouser, Mansfield, TX | 140 |
| 0-50 V power supply | 29612 PS (1) | MPJA, West Palm Beach, FL | 100 |
| Laser cut & engraving | Mask holder (1), Lens mount (1) | Ponoko, Oakland, CA | 100 |
| Analog relay timer | H3CR-A8-AC100-240 (1) | Mouser, Mansfield, TX | 70 |
| T-slotted aluminum frame | 1010 (160 in) | Knotts Co, Berkeley Heights, NJ | 40 |
| **Total** | | | **7,420** |

A detailed cost breakdown of all the main components of the CBMA is provided in **Table 2.** The major cost comes from the use of translation and rotation stages, which provide the high precision, stable, and smooth motion of the mask, wafer, and microscopes, thus allowing for a fine and accurate alignment. Other components include a collimated UV light source, custom-machined wafer chuck and mask holder, digital microscopes, and other minor mechanical and electronic components. All of these components are discussed in detail in the subsequent sections of this manuscript. The entire unit can be mounted on an optical breadboard with imperial threaded holes in order to dampen vibrations.

### 2.1. Light Source
The LED-based light source for the CBMA is reproduced from a system published by Erickstad et al.[11] This system was selected because it offers several advantages such as excellent illumination uniformity, low cost, short response time, and stable and low power consumption. Hence, the LED light source is significantly cheaper and more energy efficient when compared to a standard UV mercury lamp, which requires significant capital investment and annual maintenance, costs.

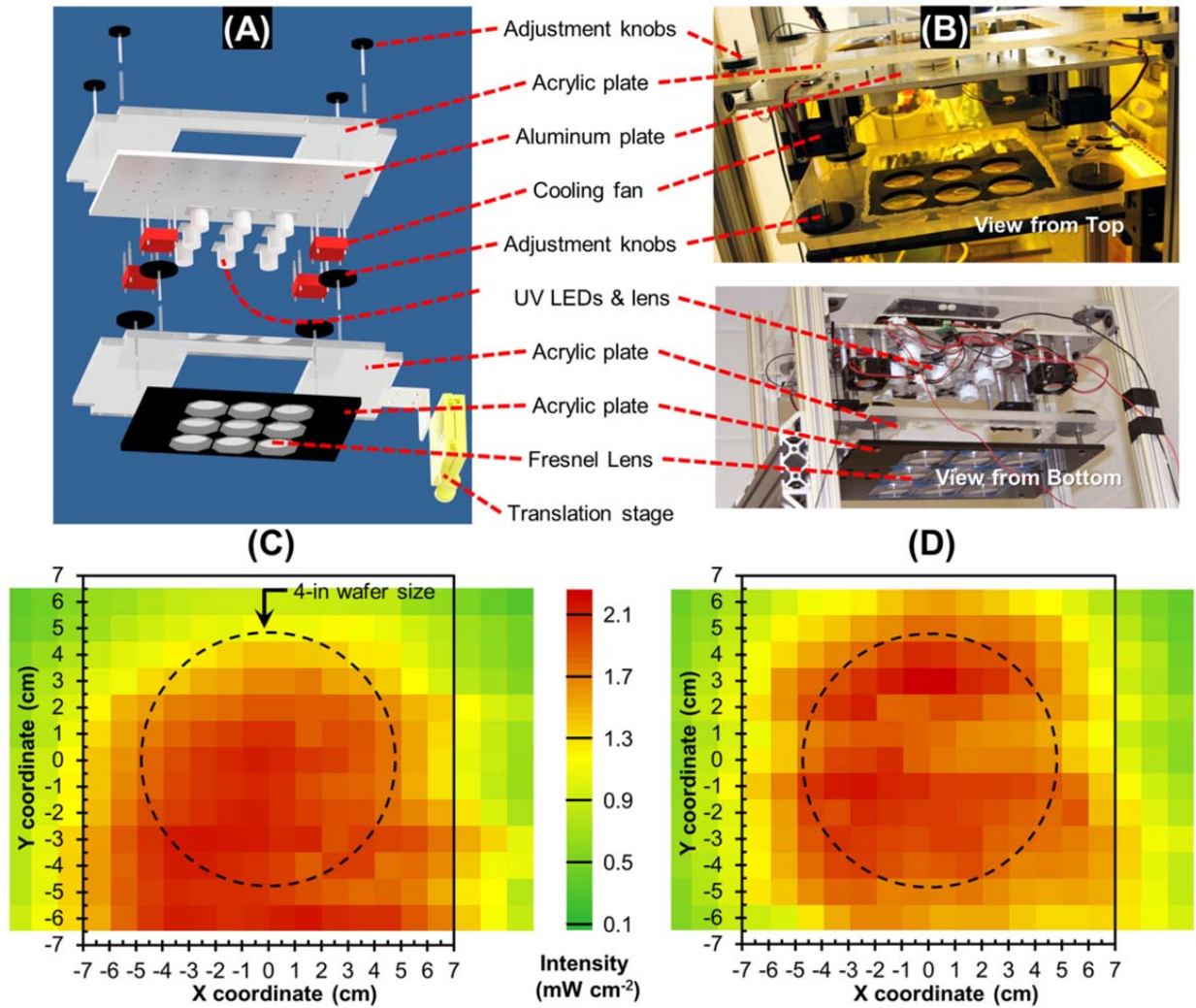

**Figure 1**. (A) An "exploded" view of the custom UV light source constructed from a squared array of 9 LEDs. Compared to original design,[11] the tip/tilt adjustment is added in order to further to aid the accurate alignment of the light source components. (B) Photograph of the light source assembly viewed from the top and from the bottom of the mask aligner. (C) & (D) Heat maps showing the distribution of light intensity before and after LED alignment, respectively. The dashed circles represent the location of a 4-in Si wafer, which is aligned co-centrically with the LED array.

The complete technical details of the LED light source have been discussed elsewhere.[11] For the sake of brevity, we describe some its main components, and the modifications we made to the original design. Typically, the light source consists of a 3x3 square array of 9 LEDs mounted on an aluminum plate, 50 mm apart from each other. The LEDs were soldered to 26G electric wires using low temperature solder cores (Indium Corporation, Clinton, NY). A UV-collecting lens was glued to each LED to reduce the divergence angle of the emitted light from 80° to 12°. The divergence angle was further brought down to ~4.7° using a squared array of credit card-sized plastic Fresnel lenses (with the same pitch as the LEDs). Light passing through the lens array is projected down to create an overlapping area of uniform illumination at the wafer plane. The plane is located 906 mm away from the aluminum plate that serves as a base for the LEDs.

In the original design, the light source was fixed to a ceiling of a room. This does not provide flexibility to the system, especially when it needs to be moved around the lab space. Instead, we mounted the light source on a supporting frame made from T-slotted aluminum 80/20 extrusions (Knotts Co, Berkeley Heights, NJ). This way, the frame and the light source can be moved together easily. We further modified the system by introducing a tip/tilt translation to the light source using an acrylic plate with adjustment knobs (**Figure 1**A, B). This adjustment ensures the perpendicularity of the light beam. The tip/tilt translation and a linear translation were also added to the Fresnel lens plate to make sure the lenses were co-aligned with the light source. Five cooling fans were provided to dissipate heat generated from long-exposure LEDs. However, they were usually turned off to avoid vibration when working with patterns of small feature sizes, since they require a short exposure time. It was found that doing this would not significantly affect the performance of the CBMA.

The adjustment of the illumination is demonstrated in Figure 1C, D. Here, it can be seen that without the adjustment, the illumination was not well centered, and is thus unable to provide a uniform photoresist exposure over a wafer surface (Figure 1C). However, by tuning the tip/tilt adjustment knobs as well as the translation stage, we were able to center the light, thus providing a more uniform illumination over the wafer area (Figure 1D). In order to generate Figure 1C and D, the spatial distribution of the light intensity at the theoretical wafer plane (i.e., ~906 mm from the light source) was measured using a Traceable UV meter (#06-662-65, Fisher, Waltham, MA) at 1-cm increment in the X and Y directions. The light source was supplied with 34-V, 0.6-A power, and the heat maps were generated using the "Color Scales" function of Microsoft Excel (Microsoft, Belleville, WA).

## 2.2. Vacuum Wafer Chuck and Mask Holder

In order to securely fasten the wafer and the photomask during alignment and exposure, an aluminum chuck (machined by Emachineshop, Mahwah, NJ) and an acrylic holder (laser cut and engraved by Ponoko, Oakland, CA) were designed (**Figure 2**). Both the wafer chuck and the mask holder operate via vacuum. The wafer chuck was patterned with 0.8-mm wide, 1-mm deep concentric troughs, which help to hold the wafer tightly and minimize any vibration due to rotation (Figure 2B). The chuck is compatible with a wafer size of 4 in or smaller. The vacuum was supplied to the innermost trough through 4 vacuum holes, which were connected to a 15-Torr house vacuum via a bearing adapter (machined by Zera Development Co, Santa Clara, CA) attached to the bottom of the chuck (Figure 2D). This setup allows the chuck to be rotated freely in 360°, thus offering a high-flexibility alignment. The holder for the photomask was laser-cut from a 10-mm thick acrylic plate in order to create a window with 4 in square aperture for the UV light. Additionally, it was engraved with a 1-mm deep, 3.25-mm wide vacuum channel for holding the mask. The holder was also connected to the house vacuum via tubing, as shown in Figure 2C. Once the vacuum was activated, the chuck could firmly hold a 5-inch square 90 mil-thick quartz mask (Photomaskportal, Richardson, TX). Both the wafer chuck and the mask holder were levelled using a bull's eye level to ensure that the mask and the wafer were in full contact. The 3D drawings for the wafer chuck and mask holder assemblies can be found in the supplemental materials.

## 2.3. Alignment Microscopes

The registration of multiple masked patterns on a wafer requires placement of "alignment" marks on the masks and "reference" marks on the wafer. This process relies heavily on the visibility of the micron-size marks, often located diametrically opposite to each other on the wafer / mask.

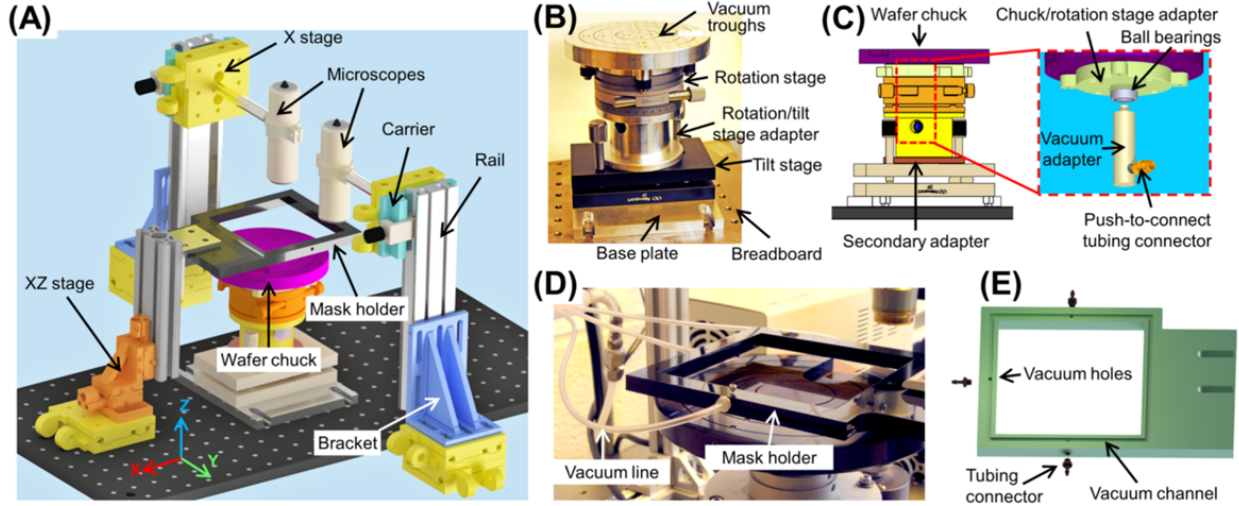

**Figure 2.** (A) A 3D illustration view showing the design of the alignment assembly comprised of a wafer chuck, mask holder, microscopes, and high-precision positioning stages. (B) A photograph of a wafer-mounting assembly showing the wafer chuck placed on top of a rotational and tip/tilt translational station. (C) An illustration showing different parts of vacuum adapters that help to connect the wafer chuck to the house vacuum. (D) A photograph showing an acrylic mask holder connected to the house vacuum via tubing. (E) A 3D view of the mask holder showing a vacuum channel engraved on the holder's surface, and holes supplying vacuum to the channel.

Herein, we used two digital microscopes (AD4113T, Dino-lite Digital Microscope, Taiwan) positioned on both sides of the CBMA to provide microscopic views of the marks (Figure 2A). These microscopes were used since they are cheap, compact, and simple to operate. Moreover, they can be connected to a laptop via a Universal Serial Bus (USB) interface, and controlled using the Dino Capture 2.0 software provided by the manufacturer. During alignment, both microscopes were used simultaneously to acquire a live view of the marks at various magnifications (**Figure S1**). The UV content of the microscopes' light sources is negligible, as measured by a UV meter; however, to ensure the photoresist was completely intact during alignment, the microscopes were covered with amber UV filter films (#F007-006, UVPS, Chicago, IL), which absorbed any UV radiation emanated from the microscopes' light sources.

## 2.4. Kinematic Mask Alignment System

In order to ensure of a robust mask-to-wafer alignment with a high fidelity, high-precision positioning stages were used. The wafer chuck was mounted on a rotary stage (#UTR80, Newport) and a tip/tilt stage (#39, Newport), which offer high sensitivities of 0.001° and 0.002°, respectively (see Figure 2B). The mask holder was mounted on a translation complex, which consisted of a XZ translation stage (#461-XZ-M, Newport) and a Y translation stage (#423, Newport) combined. This allows the mask to travel in all three directions, with travel distances of 1 inch in X and Z, and 2 inches in Y (see Figure 2A). Digital microscopes were also mounted on linear translation stages (#423-MIC, Newport, Irvine, CA) configured in XYZ directions, thus allowing the microscopes to be accurately positioned and focused in a highly repetitive manner. Fast and coarse vertical motion of the microscopes was facilitated using tail optical rail and carrier (10R300 and 20C, respectively; Optics Focus Instruments Co, China).

## 2.5. Electronics

A bench-top variable 0-50 V power supply (#29612 PS, MPJA, West Palm Beach, FL) was used to power the light source using a voltage setting of 34 V and a current of 0.6 A. This provides an average illumination power of ~1.85-2.00 mW/cm$^2$ at the wafer's surface (see Figure 1C, D). The exposure time was controlled by a mechanical relay timer from Omron (#H3CR-A8-AC100-240, Mouser, Mansfield, TX) which automatically disconnects the light source from the power supply when the set time is reached. The cooling fans (#SY124020L, Scythe Co., Germany) were powered by a 12V AC power supply (#1670, Current USA, Vista, CA).

## 2.6. Supporting Frame

The frame of the CBMA was constructed from T-slotted 8020 aluminum extrusions (Knotts Co, Berkeley Heights, NJ) which offers stable mounting of the hardware components, and at the same time provides for the light-weight nature of the CBMA (Figure S1). The light source was attached directly to the frame, while the wafer chuck, the mask holder, and the microscopes were not connected to it. This frame can be conveniently mounted on top of an optical table with imperial threaded holes. Alternatively, it can also be used as a standalone setup, without the optical table. However, one is recommended in order to reduce fabrication defects caused by vibration. This setup offers the flexibility often desired in laboratories with a limited space.

## 3. Fabrication Procedure

The fabrication of a master mold of a microfluidic device (shown in **Figure 6** and discussed in detail in the "Mask Aligner Application" section) with two different heights is used to illustrate a typical alignment procedure for the CBMA. Generally, this type of fabrication follows a standard photolithography technique, which includes three basic elements: spin coating, UV exposure, and development. In order to generate a master mold with two different heights, we used three photomasks: a "reference" mask containing only reference marks to be imprinted on the wafer, and two "device" masks containing templates for the different photoresist heights. These masks can be either chrome coated on a glass substrate or printed on transparencies at a high resolution (> 10,000 dots per inch). Even though the transparency masks are not likely to offer feature sizes smaller than 10 μm, they are inexpensive and suitable for most standard microfluidic applications. Therefore, we went with this option.

The transparency masks were taped onto 5-in square bare quartz substrates (90-mil thickness) (Photomaskportal, TX) with the printed side facing outward, using a ¼-in polyimide film tape (#5413, 3M, Mapplewood, MN). A 3-step procedure for the fabrication of a multi-height master is outlined in **Figure 3**. In the *first step* of this procedure, the two reference mark arrays were fabricated diametrically opposite to each other on a 4-in Si wafer (#1196, University wafer, South Boston, MA), using the "reference" mask. The marks were 70-80 mm apart from each other. Positive photoresist (# AZ P4620, Microchem, Westborough, MA) was diluted prior to spin-coating with an edge bead removal solvent (#EBR PG, MicroChem). A resist:EBR ratio of 1:4 was used to achieve a thin layer of 5 μm. This photoresist was selected due to its amber shade when developed, which provides alignment marks that are highly contrasted relative to the color of the wafer. The alignment accuracy is significantly enhanced as a result.

In the second step of the procedure in Figure 3, the 20 μm-high feature of the master mold was generated from a negative photoresist SU-8 2015 (Microchem). The negative photoresist was spin-coated on the "marked" wafer generated in step 1. After the coating, the first layer "device" mask is aligned with the wafer's reference masks using the alignment procedure discussed in the subsequent "Mask Alignment Procedure" section of this manuscript. After alignment, the wafer was exposed to UV (140 mJ), and developed to create a 20-μm high resist pattern. The *third step* of the procedure in Figure 3 is needed in order to generate the 60-μm high feature of the master mold. It essentially repeats the *second step*, but with a different "device" mask. Photoresist SU-8 2035 (Microchem) was again spin-coated on the same wafer to achieve a 60-μm thick film. This was then followed by alignment of the second layer "device" mask using the CBMA. The resist was then exposed to UV (210 mJ) and developed to attain the final master mold.

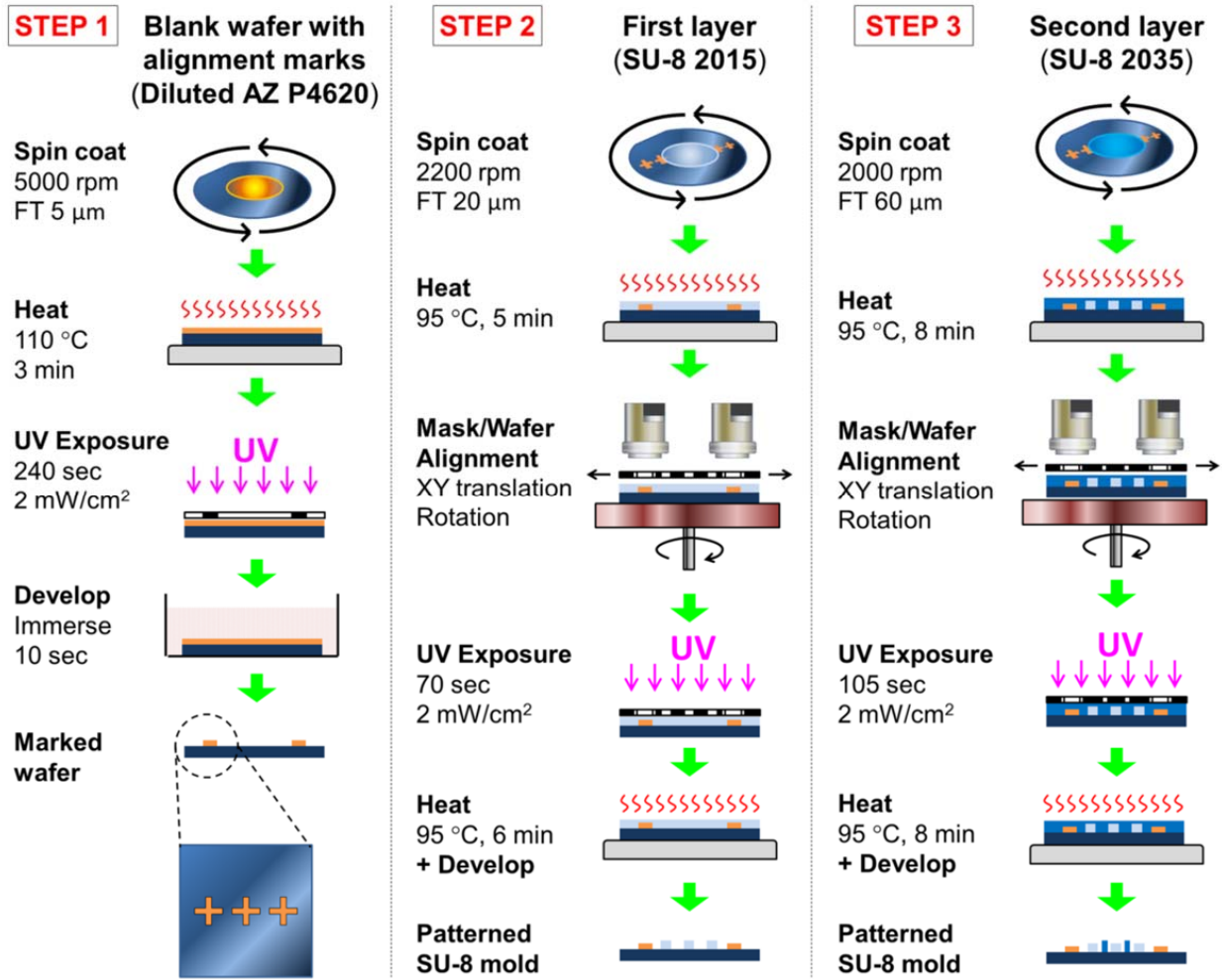

**Figure 3**. Fabrication of a multi-height master mold on a 4-in Si wafer using the CBMA in a 3-step procedure. In step 1, reference marks are imprinted on the wafer using a diluted positive photoresist (#AZ P4620, Microchem) via a "reference" mask. In step 2, a 20-μm high SU-8 photoresist layer is patterned on the wafer, which is aligned with the mask using the CBMA. In step 3, a 60-μm thick SU-8 layer is patterned on the same wafer following a similar procedure mentioned in step 2, but with a different mask. Abbreviations: FT = Film Thickness.

## 4. Mask Alignment Procedure

A detailed aligning procedure is illustrated in **Figure 4**A. First, a "device" photomask is placed on top of a vacuum wafer chuck with the pattern side facing downward. The mask holder is lowered in the Z direction by a XYZ translation stage, until it is in contact with the mask. Then vacuum is applied in order to seal the mask to the mask holder. Once the mask is securely attached to the holder, the assembly is raised up in order to vacuum-seal a marked wafer (described in Step 1 of Figure 3) on to the wafer chuck. The mask is then brought in close proximity to the wafer (i.e. less than 100 µm, assuming the resist film thickness is smaller than that). Note that the distance between the mask and the wafer can be judged by observing the reflection of the mask on the wafer surface. Excessively close proximity is not necessary, as it would restrict the movement of the mask and the wafer during alignment.

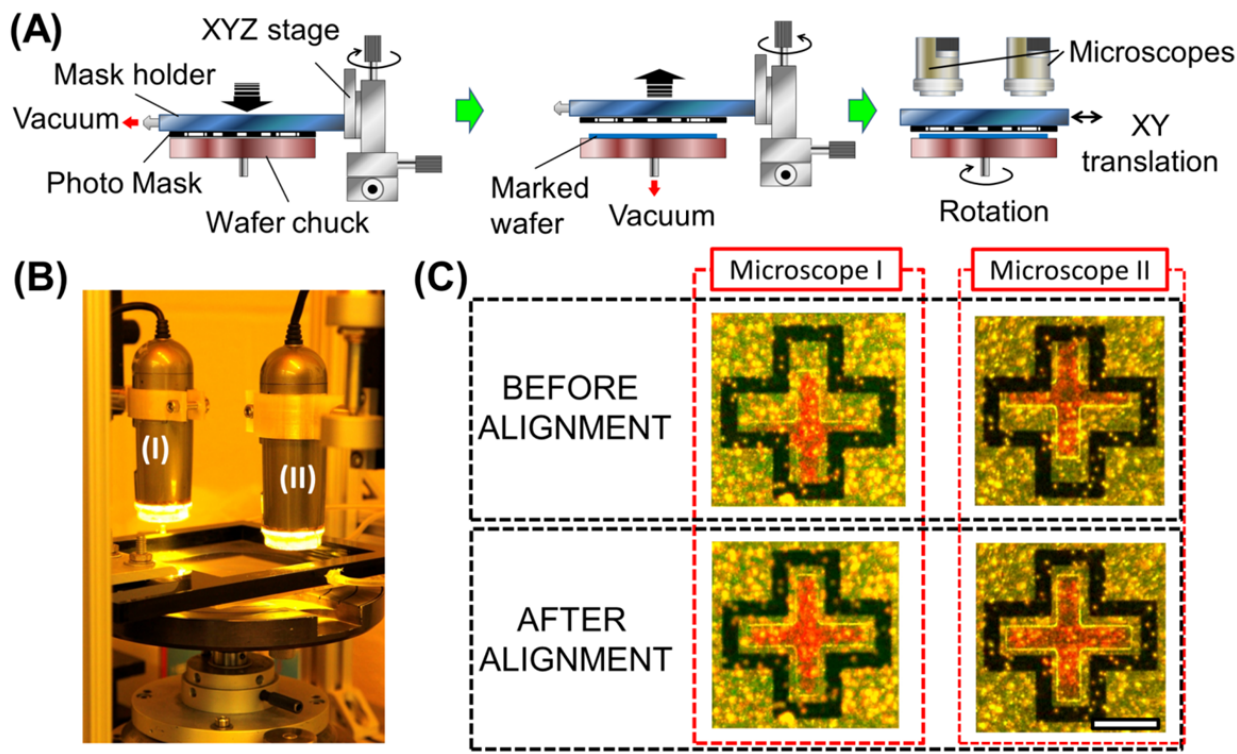

**Figure 4**. (A) Diagram showing a step-wise alignment procedure using the CBMA. (B) Two alignment microscopes in operation. (C) Visualization of the fine alignment process captured by the alignment microscopes at 245x. Black hollow and red solid crosses are the mask "alignment" and the wafer "reference" marks, respectively. Scale bar is 200 µm.

The coarse alignment begins by focusing two digital microscopes on the two alignment mark regions on the "device" masks, as shown in Figure 4B. The reference marks on the wafer are brought into the microscopes' views using the translation and rotation stages. This is first done at a 30x magnification of the two microscopes in order to achieve coarse alignment (**Figure S2**A). Fine alignment then follows at a 245x magnification. At this time, the marks viewed by Microscope I should be symmetrically opposite to those viewed by Microscope II with respect to the center of the wafer. This helps to minimize the rotational error of the alignment. In addition, it should also be noted that at the 245x magnification, the Dino-Lite microscopes have a working distance of ~ 10 mm. This allows both the wafer's reference marks and the mask's alignment marks to stay in focus at the same time (Figure 4C), which also greatly enhances the alignment accuracy.

The fine adjustment is first started by bringing the two masks' alignment marks to the same Y position on the observation monitor (see coordinate axis in Figure S2B). Rotation of the wafer is then performed in order to bring the wafer reference marks to the same Y position on the view screen. After that, the rotation stage is locked to restrict the rotation movement of the wafer. This is followed by bringing the mask closer to the wafer by adjusting the Z stage attached to the mask holder. The actual proximity depends on the operation type that is most appropriate for the user's purpose: either contact-free or hard contact. Following that, the mask is then translated in the XY plane in order to center the wafer reference marks inside of the mask's alignment marks, as shown in Figure 4C. Once the alignment is finished, the two microscopes are moved out of the way, and the light source is turned "on" in order to start the exposure.

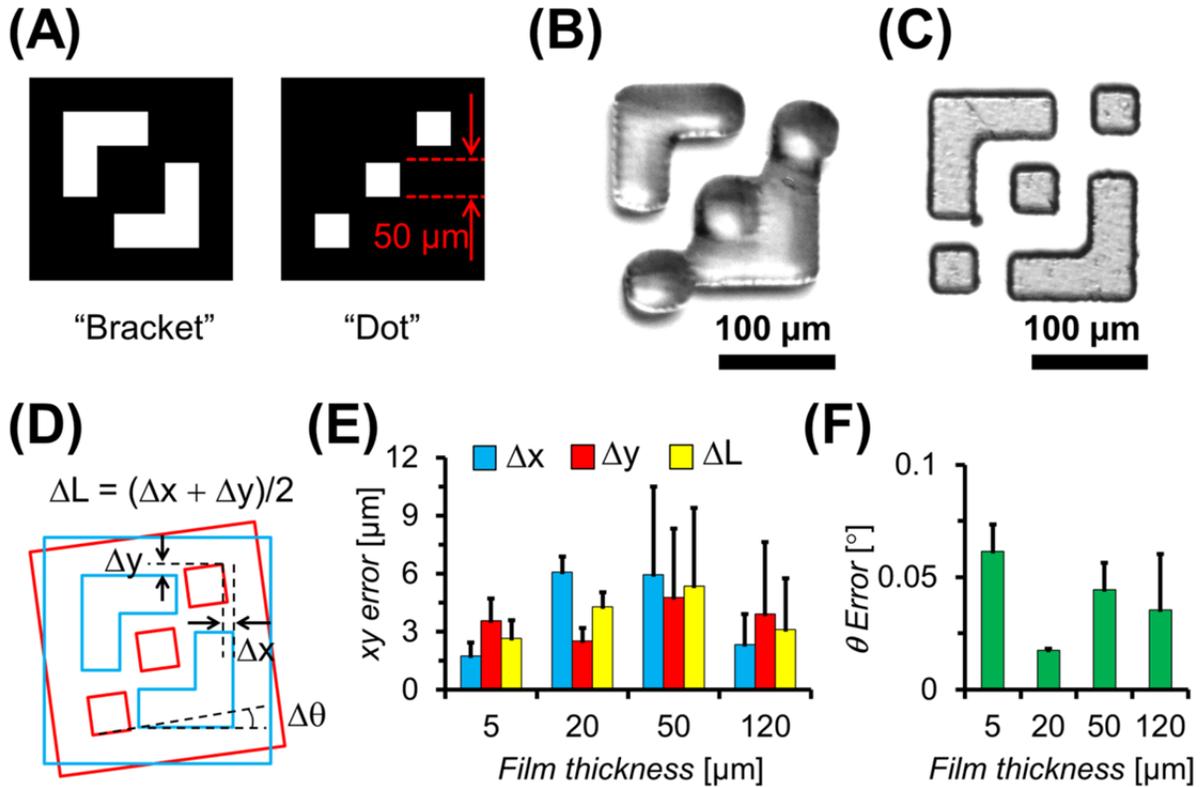

**Figure 5**. (A) Two photomasks containing different alignment patterns, "bracket" and "dot", were used to generate a photoresist profile for the alignment accuracy measurement. (B) Micrograph of a defective SU-8 print caused by misalignment of the two patterns. (C) Micrograph of a SU-print obtained from well-aligned patterns. (D) Illustration of how translational x-y and rotational θ error values are determined. (E) & (F) Alignment errors quantitatively determined from the SU-8 prints with respect to the photoresist's thickness.

## 5. Characterization of Mask Alignment Accuracy

The accuracy of the alignment was quantitatively determined from SU-8 prints which are constructed using two different mask patterns, as shown in **Figure 5**A. The masks with feature size of 50 μm were printed on plastic transparencies at 10,160 DPI (Fineline Imaging, Colorado Springs, CO). The first mask contains "bracket" patterns and the second mask has "dot" patterns. SU-8 resist films of various thicknesses, ranging from 5 μm to 120 μm, were coated on a 4-in Si wafer containing positive photoresist alignment marks. The "bracket" mask was then aligned with the wafer following the protocol mentioned in the previous section. The photoresist layer was then exposed to UV radiation for a duration of 40 to 120 seconds, depending on the film's thickness. After that, the "bracket" mask was replaced by the "dot" mask, and followed by alignment. The resists were then exposed at the same UV dose, baked at 95 °C for 5-10 min, and developed to generate the measurement patterns (see Figure 5B, C). Images of the patterns were acquired using a CCD camera Guppy Pro F-146 (Allied Vision, Germany) coupled to a reflected microscope Reichert Zetopan Trinocular (Austria). The acquisition was performed using a 10X objective (PLN10X, Olympus, Japan). Measurement of the patterns was conducted using open source ImageJ software.[13]

Upon successful alignment, the final SU-8 replica should contain the "dot" pattern centered inside of the "bracket" pattern, without offsets in any direction. Severe misalignment leads to the overlapping of the "dot" on the "bracket", as shown Figure 5B. In contrast, careful alignment results in a well-defined print that looks closer to what is expected from the superposition of the two masks (see Figure 5C). The aligned patterns were then used for the characterization of the alignment accuracy. Figure 5D demonstrates how the translational and rotational alignment errors are determined quantitatively.[12] Namely, the translational errors $\Delta x$ and $\Delta y$ were calculated from the offset of the "dot" pattern with respect to the "bracket" pattern in the x and y directions, respectively. In addition, the mean translational error $\Delta L$ was calculated by averaging the translational errors in the two individual directions. On the other hand, the rotational error $\Delta\theta$ was determined from the angular amount that the former pattern was offset from the latter one. In practice, errorless alignment (i.e., zero delta values), is not possible even with a state-of-the art mask aligner, which can achieve sub-micron alignment accuracy.

In essence, the alignment accuracy depends strongly on the user's ability to observe the alignment marks on the mask and the wafer. This, in turn, is dictated by the resolution and by the image quality of the microscopes used for the alignment. Thus, in order to attain optimal alignment accuracy, the digital microscopes were used at their maximum magnification of 245X (corresponding to a resolution of 1.25 μm per pixel). However, other factors such as human error and film thickness[14] may contribute to the imperfection of the alignment. Therefore, the accuracy of 1.25 μm is more or less theoretical for our system, especially considering that the resolution of the stages is also ~1μm.

It has been reported elsewhere that the film thickness in particular affects the alignment accuracy by increasing the distance between the mask and the wafer.[14] This makes it progressively more difficult to keep both the mask's marks and the wafer's within the microscope's depth of field. Consequently, the inability to focus on both of the marks simultaneously creates more room for error. We therefore characterized how the alignment accuracy of our CBMA varies with respect to the different photoresist thicknesses: ranging from 5 to 120 μm. Given that the alignment accuracy is also dependent on the operator's skill and experience, we performed each of the thickness experiments in triplicate in order to ensure that the human error was minimized.

Surprisingly, both the translational and the rotational accuracies were not found to be highly dependent on the film's thickness (Figure 5E, F); however, the thinner ones resulted in a higher consistency (Figure 5E). With film thickness of 50 μm or thinner, the translational errors appears to increase with the resist's thickness. Typically, the 5-μm film thickness resulted in average translational and rotational errors of 2.96 ± 0.92 μm and 0.062 ± 0.012°, respectively. Alignment errors of 20 and 50-μm thick prints were 4.29 ± 0.74 μm, 0.018 ± 0.0007° and 5.36 ± 4.04 μm, 0.045 ± 0.012° respectively. On the other hand, the rotational errors did not exhibit any distinguishable pattern, with values ranging well below 0.1° for all thickness values. This is likely due to the high resolution of the rotation stage, which is about an order of magnitude higher than the alignment accuracy of the translational stages. Finally, to our surprise we were able to achieve a high accuracy with the 120-μm thick film (3.12 ± 2.65 μm, 0.036 ± 0.025°), suggesting that film's thickness is not the only factor that hampers the alignment. For example, other factors like vibration, glare, unevenness in the transparency mask, and others could have also contributed to the variation in the alignment results.

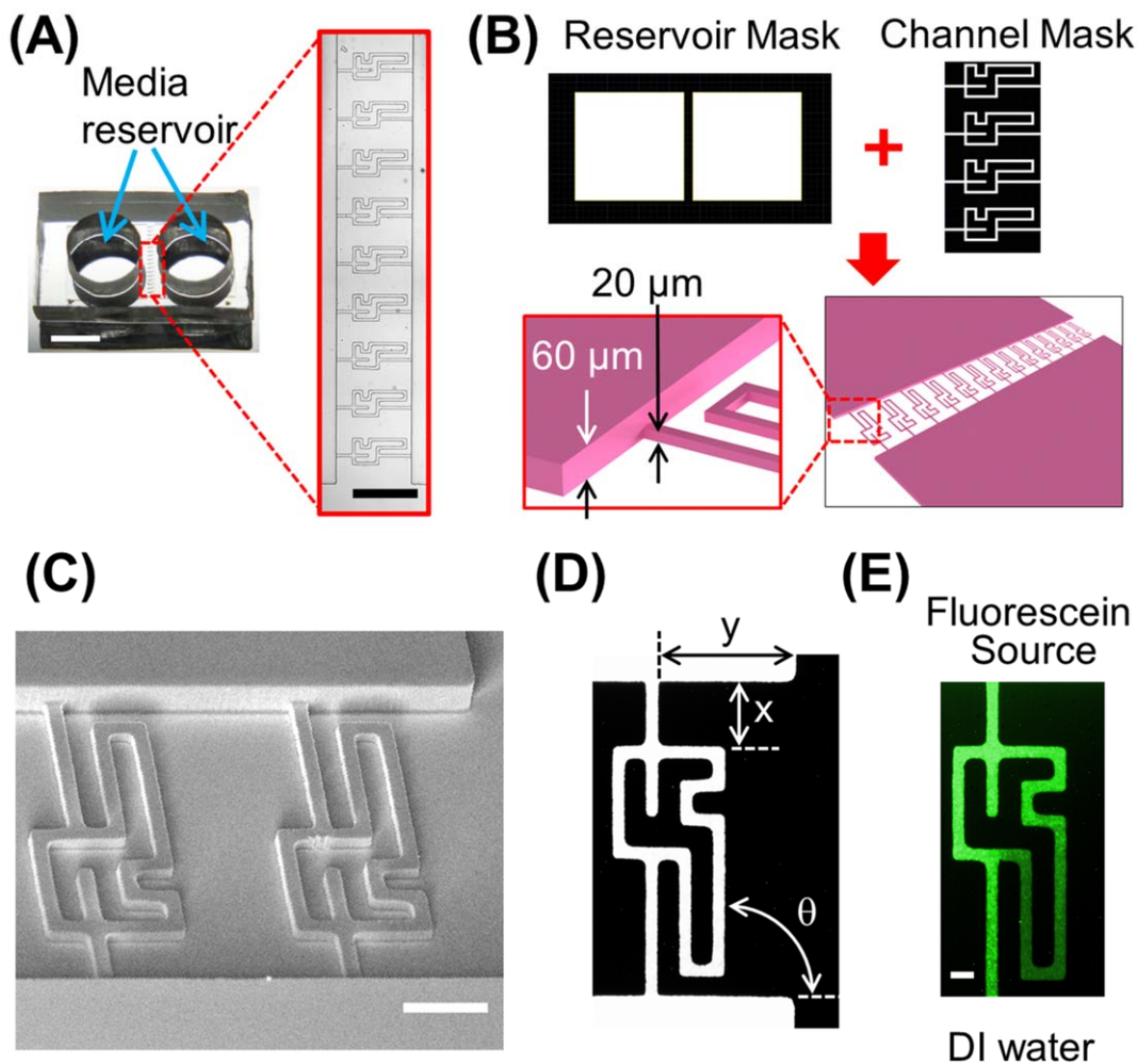

**Figure 6**. (A) A photograph of the chemotaxis microfluidic device used as a demonstrative application of our CBMA. The device consists of an array of 15 x 20 (W x H) μm channels connecting two media reservoirs of 60-μm in height. Inset is a bright-field microscopy view showing how the microfluidic channels connect the two media reservoirs. Scale bar is 4 mm for the photograph and 400 μm for the inset. (B) Two photolithography masks used to fabricate the master mold for the device: one for the reservoir and one for the channel. Bottom is a 3D drawing of the resulting mold. (C) SEM image of the fabricated two-height SU-8 master. Scale bar is 100 μm (D) Interpretation of how the alignment of the device is characterized. The quantities Δx, Δy, and Δθ are calculated from the difference in x, y, and θ between the device and the reference mask. (E) Fluorescent microscopy image of fluorescein gradient formed inside the micro channel of a PDMS device. Scale bar is 50 μm.

**6. Mask Aligner Application**
In order to demonstrate an application of our CBMA, we fabricated a SU-8 master mold for a PDMS device similar to ones commonly used to study cell chemotaxis in response to a concentration gradient of a chemoattractant (see **Figure 6**A).[4, 15-17]. The device consists of an array of 20 μm-high diffusion "maze" channels that connect two identical reservoirs, each of which are 60 μm in height. In the chemotaxis application, one of the reservoirs is meant to hold the cells, while the other typically serves a source of the chemoattractant. In order to generate the device, a master of two different heights (20 and 60 μm) was fabricated from SU-8 photoresist using two different masks: one for the reservoir and one for the diffusion channel (see Figure 6B). The distance between the two reservoirs was 462 μm and the distance between the two ends of the channel was 482 μm. Since, carrying out experiments in parallel requires that the mazes are identical, it is critical to align the two mask patterns accurately. Moreover, any mismatch between the diffusion channel and the reservoir would result in the malfunction of the device. Scanning electron microscopy (SEM) image (Figure 6C) shows that a two-thickness master was successfully generated. SEM was acquired at 45° tilt angle using Field Emission Scanning Electron Microscope (FESEM, LEO 1530VP, Zyvex, Richardson, TX). Prior to imaging, the sample was sputter coated with 5-nm thin gold layer using a sputter coater (EMS150T ES, Quorumtech, Lewes, UK).

The alignment accuracy of the device was characterized in the x, y, and θ directions (Figure 6D), using the same procedure as in Figure 5D. The quantities $\Delta x$ and $\Delta y$ were determined to be ± 3.9 μm, while $\Delta \theta$ was 0.02°. This result suggests a successful alignment of the device.

We further investigated the performance of the device by testing chemical gradient generation within the diffusion channel using fluorescence tracer Alexa Fluor 488 (Thermofisher, Waltham, MA). The top of the reservoirs was punctured by a 5-mm biopsy punch in order to form a circular inlet port for the liquid media input. The fluorescent solution (Alexa Fluor 488 (1 μg ml$^{-1}$) in deionized water) was added to one of the reservoirs, while deionized water was added to the other reservoir. Figure 6E show that the resulting concentration profile of the tracer established inside the channels, between the source and the sink of the fluorescein. This result suggests that the device functioned successfully without any defects (i.e. no mismatch between the channels and the reservoirs) and was able to generate a stable gradient, which is suitable for a chemotaxis study of cells. Fluorescence microscopic images were acquired using an inverted microscope (IX83, Olympus, Japan) coupled to a 488-nm laser light source.

In a second application, we use our CBMA to fabricate something even more complex: a 3-height replica of an addressable stencil device, which was inspired by a work published by others.[6] This type of device is commonly used to deliver varied concentrations of chemicals to different "addresses" in the chip. Typically, the stencil device is comprised of three PDMS layers: a "control" layer, a "flow" layer, and a "display" layer. However, since the "control" and the "display" layers require just a single-height master mold (which could be easily fabricated using just a UV lamp without the need for a mask aligner), we focus exclusively on the fabrication of the 3-height "flow" layer master-mold. This fabrication includes the use of 3 different mask patterns: a 24-μm high main flow channel to transport liquid from a media source, a 25-μm high bypass channel, and a 90-μm high hole pattern to deliver media to the "display" layer underneath (**Figure 7**A).

In order to generate this master mold, a 3-exposure process was conducted using 3 different photomasks, each of which corresponds to a featured pattern. Firstly, the flow layers was fabricated by double coating the positive photoresist AZ P4620 to obtain 24-μm thick film on an "reference" mark-containing 4-in Si wafer, following a 3-min soft bake at 110 °C. Then the wafer was aligned with a photomask containing the pattern for the flow channel, and exposed to UV (2 mW cm$^{-2}$) for 960 seconds using the CBMA. The resists were developed and heated to 150 °C on a hot plate for 20 min to generate a round channel profile, as well as for improving the adhesion of the resist to the wafer for subsequent fabrication. After that, the wafer was coated with a 25-µm thick layer of the negative photoresist SU-8 2015, aligned with the second mask containing the pattern of the bypass channels, and exposed to 75 seconds of UV (2 mW cm$^{-2}$). It was then developed and heated on a hot plate at 150 °C for 5 min. Next, the wafer was coated with 90-µm thick SU-8 2035, aligned with the third mask containing the pattern for the holes, and exposed to UV (2 mW cm$^{-2}$) for 110 seconds. Finally, the wafer was developed, and hard baked at 150 °C for 5 min to generate the final master. The master was then examined using SEM. The SEM images reveal that the master was successfully generated, with all the patterns were well aligned with each other (Figure **7**B).

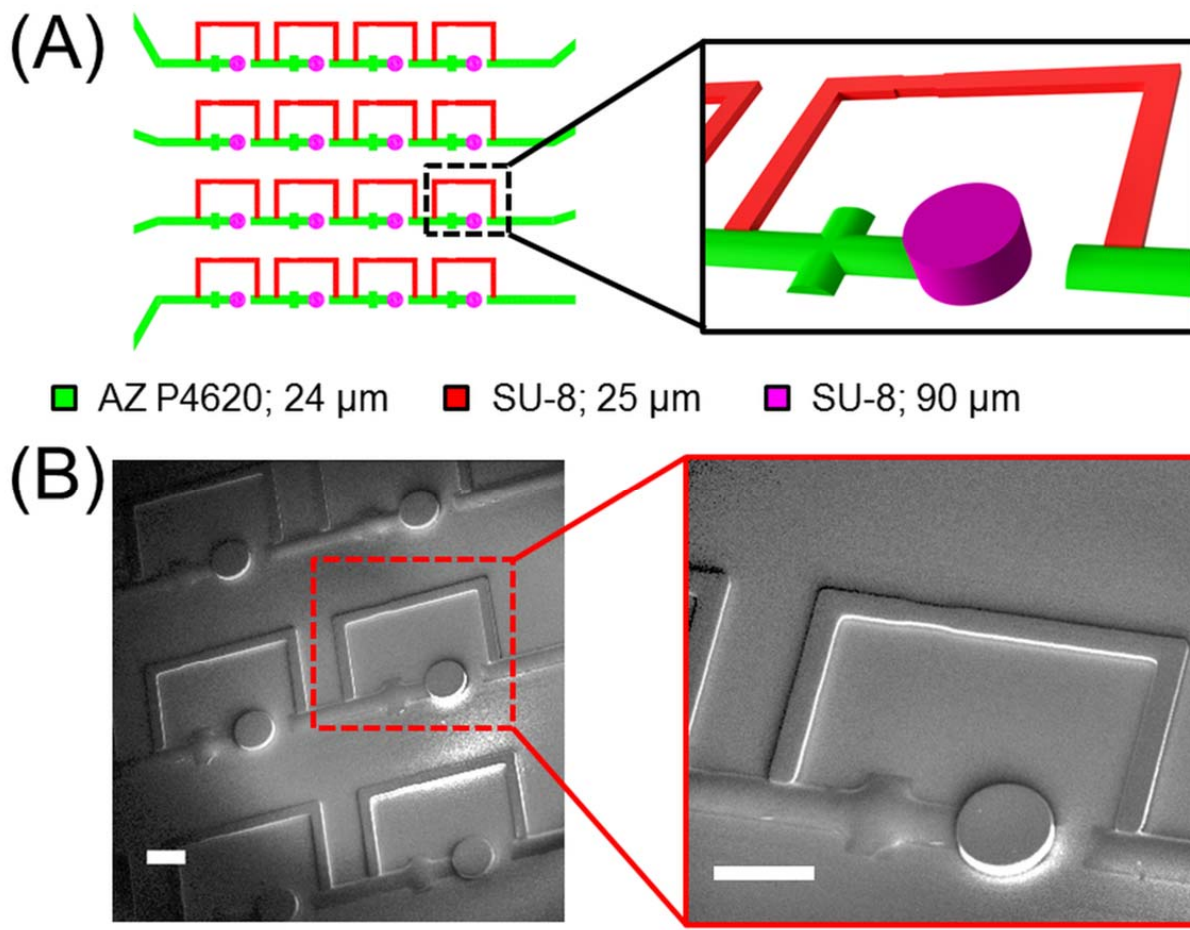

**Figure 7**. (A) Design of a master mold for the "flow" layer of an addressable stencil device that consists of three photoresist patterns of various heights: 24-µm high main flow channel made out of positive photoresist AZ P4620 (green), 25-µm high bypass channel made out of negative photoresist SU-8 (red), and 90 µm-high delivery hole made out of SU-8 (pink). Colors indicate individual photoresist patterns achieved via different photomasks. (B) SEM images showing different feature heights of the resulting master mold. Scale bars are 200 µm.

## 7. Conclusion & Future Work

In this paper, we demonstrated a cost-effective, lightweight, and accurate custom-built CBMA for the fabrication of multi-height photoresist patterns which are subsequently used as master molds for microfluidic applications. The CBMA was approximately tenfold cheaper than analogous commercial systems, thus making it suitable for medium and small laboratories in need of in-house microfluidic device fabrication. Owing to its lightweight and space-saving feature, the CBMA can be integrated into any lab space as a bench top unit. In addition, it is suitable for thick resist layers (> 100 μm) with an alignment accuracy of <3 μm, making it a potential replacement for the high-cost, high-maintenance commercial systems typically used in microfabrication.

Although our CBMA, is capable of achieving a wide spectrum of layered features that cover most of the microfluidics applications in its presented form, several major and minor improvements can be made to it based on the user's needs. For example, a simple enhancement can be achieved via supplementing the two alignment microscopes with a dedicated tablet PC each (simple ones can cost as low as ~$100). This can save the user time involved with connecting/disconnecting the microscopes to computers and monitors every time the CBMA is used.

Another possible upgrade is automating the alignment stages with electronic controllers. For example, the manual rotation stage can be replaced by a motorized version (#URS75BPP, Newport); while the linear translation stages can be actuated by motorized actuators (#TRA25, Newport). This can both make the resulting products more accurate and repeatable by eliminating the possibility of inaccuracy introduced by manual operation error. However, the increased price associated with the automation and the need for custom software are, of course, the trade-off with this option.

Additionally, the use of Grating Light Waves (GLV), Digital Micro-mirror Devices (DMD), Spatial Light Modulators (SLM), and Liquid Crystal Displays (LCD) as electrically controllable active photomasks has recently become increasingly common because it can significantly simplify the 3D structuring of the photoresist.[18, 19] In this case, a computer-generated pattern serves as a reconfigurable mask in the lithography system, rendering a mask alignment procedure unnecessary because the position of the photomask is always fixed. Although alignment of the wafer with the mask may still be necessary (e.g., if multiple coatings of different photoresists are required), the active-mask photolithography makes the pattern alignment precise at the exposure plane and allows for rapid fabrication of 3D microstructures. Among these, the LCD is an especially promising for high-resolution lithography, because of its small pixel size: 8.5μm vs 17μm for DMD.[18, 19] Moreover, the LCD mask enables "gray-tone" photolithography (useful for tapered devices)[20] by allowing the user to adjust the percent transmittance of each individual pixel in order to produce advanced 3D features.

Other potential design modifications could include exposing the photoresist from the bottom by defining the mask as a part of the substrate.[21] By spinning the photoresist directly on to the mask, perfect contact alignment is achieved enabling the fabrication of tall high aspect ratio structures. Finally, recent exciting developments in LED technology such as the Nano-LED single photon lithography, could potentially be implemented in our CBMA once the technology matures.[22] The advantage of this LED type is that it can significantly improve the resolution (down to molecular scale) while at the same time eliminating the need for a physical photomask if arranged in an electronically driven individually addressable array.

## Supporting Information
Supporting Information is provided as an appendix to this manuscript.

## Acknowledgements

We acknowledge the financial support from the Gustavus and Louise Pfeiffer Research Foundation and the National Science Foundation I-Corps Site Award #: 1450182. We also thank the Otto H. York Department of Chemical, Biological, and Pharmaceutical Engineering at New Jersey Institute of Technology for the help with machining custom parts for the CBMA.

# Supporting Information

**A Compact Low-Cost Low-Maintenance Open Architecture Mask Aligner for Fabrication of Multilayer Microfluidics Devices**

*Quang Long Pham, Austin Mathew, Nhat-Anh N. Tong, Roman S. Voronov\**

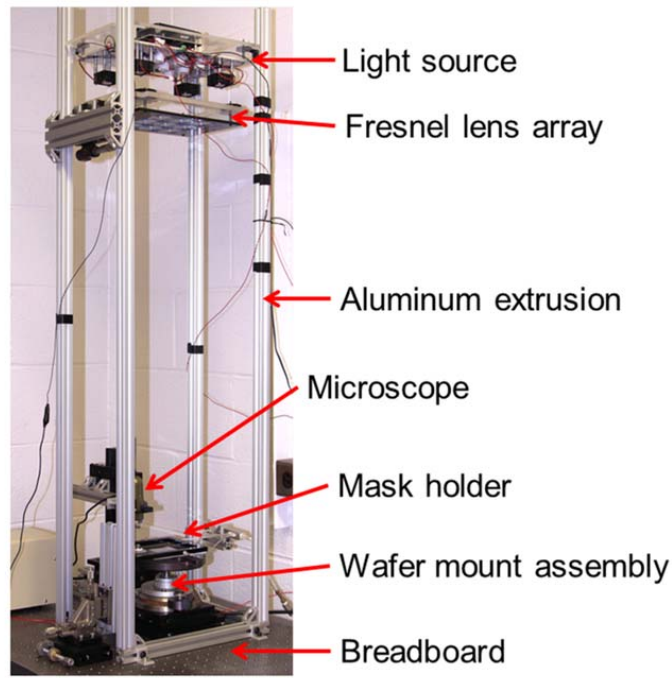

**Figure S1**. Photograph of the CBMA showing all of its major components: the light source, the microscope, the mask holder, and the wafer-chuck assembly. The light source and the Fresnel lens array are mounted on a supporting frame assembled from T-slotted 8020 aluminum extrusions.

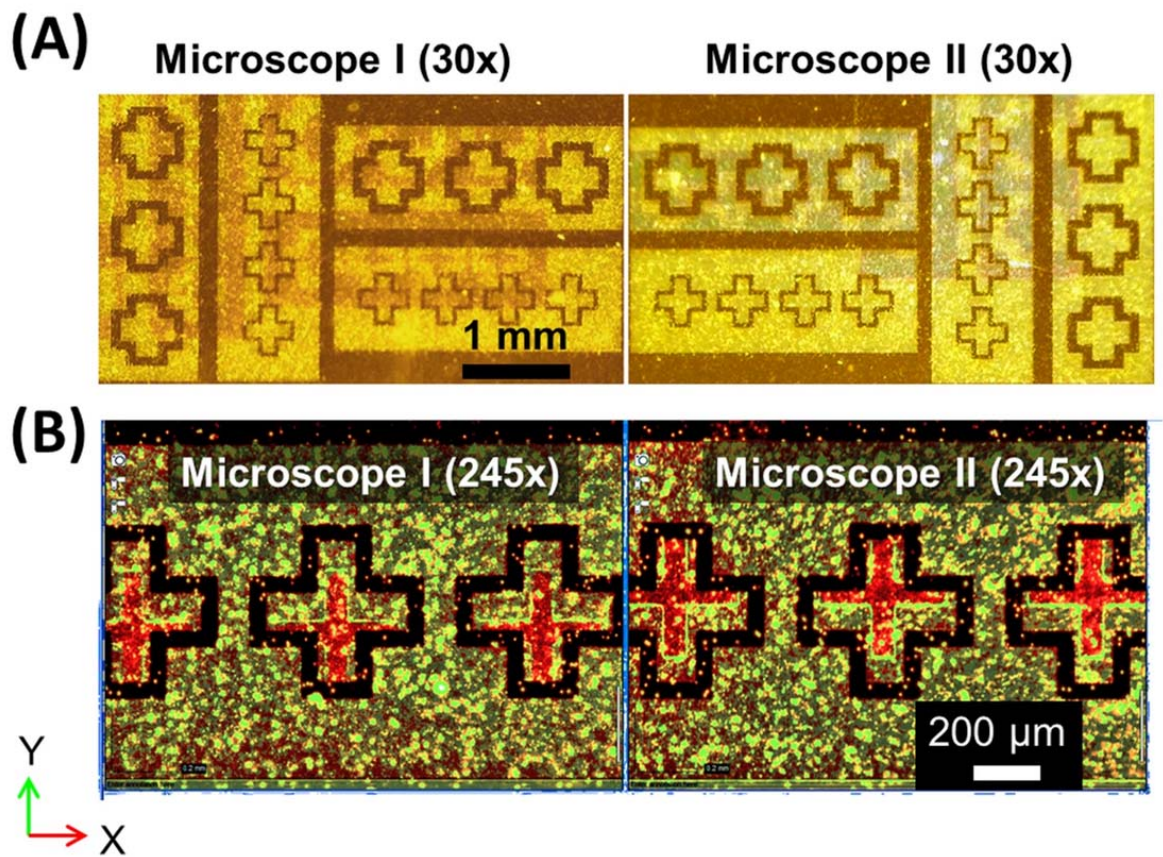

**Figure S2**. Screenshots from two monitors, showing the live-view microscopy acquisition windows corresponding to the two alignment microscopes. Alignment and reference marks are positioned at the same Y location on each respective screen. (A) Two alignment/reference mark arrays being focused upon at a 30x magnification. (B) Individual alignment/reference marks being focused upon at a 245x magnification. Alignment marks (black color) of the on the left side and on the right side of the "device" mask were brought into same Y positions on each of the view screens.

| Part Description | CAD Design | File Name | Materials | Company |
|---|---|---|---|---|
| UV LED-mounting plate | 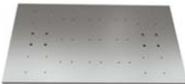 | LS-001 Aluminum Plate.ipt | Aluminum 6061 | Emachineshop, Mahwah, NJ |
| Wafer vacuum chuck | 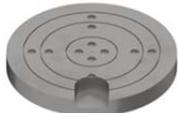 | MA-001 Wafer chuck.ipt | Aluminum 6061 | Emachineshop, Mahwah, NJ |
| Chuck-to-rotation stage adapter | 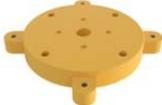 | MA-002 Chuck to rotation stage adapter.ipt | Aluminum 6061 | Zera Development, Santa Clara, CA |
| Ball bearing-to- vacuum adaptor | 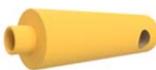 | MA-003 Ball Bearing Vacuum Adaptor.ipt | Aluminum 6061 | Zera Development, Santa Clara, CA |
| Rotation-to-tilt stage adapter | 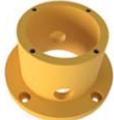 | MA-008 Rotation to Tilt Stage Adapter.ipt | Aluminum 6061 | Zera Development, Santa Clara, CA |
| Rotation-to-tilt stage secondary adapter | 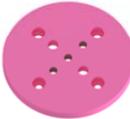 | MA-009 Rotation to tilt stage secondary adapter.ipt | Aluminum 6061 | Zera Development, Santa Clara, CA |
| Tilt stage-to-breadboard base plate | 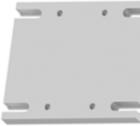 | MA-010 Tilt stage-to-breadboard base plate.ipt | Aluminum 6061 | Emachineshop, Mahwah, NJ |
| Fresnel lens holder | 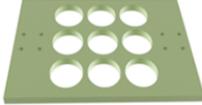 | LS-002 Acryllic Lense Holder.ipt | Acrylic | Ponoko, Oakland, CA |
| Mounting plate | 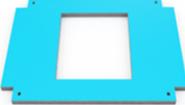 | LS-003 Acryllic Mounting Plate.ipt | Acrylic | Ponoko, Oakland, CA |
| Mask holder | 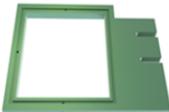 | MA-013 Mask Holder.ipt | Acrylic | Ponoko, Oakland, CA |
| Microscope holder | 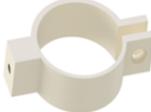 | MI-001 Dino Lite Microscope Holder.ipt | Poly (lactic acid) | In-house printed using a 3D printer (Ultimaker, UK) |

**Table S1**. List of computer-aided design (CAD) parts; 3D files are provided upon request.